\documentclass[a4paper]{article}
\usepackage{amsmath}
\usepackage{graphicx}
\usepackage[english]{babel}
\usepackage{subfig}
\usepackage{hyperref}
\hypersetup{colorlinks = true, allcolors = cyan}

\begin{document}
\title{Polyhedral design\\with blended $n$-sided interpolants}
\author{P{\'e}ter Salvi\\{\small Budapest University of Technology and Economics}}
\date{}
\maketitle
\begin{abstract}
A new parametric surface representation is proposed that interpolates
the vertices of a given closed mesh of arbitrary topology. Smoothly
connecting quadrilateral patches are created by blending local, multi-sided
quadratic interpolants. In the non-four-sided case, this requires
a special parameterization technique involving rational curves. Appropriate
handling of triangular subpatches and alternative subpatch representations
are also discussed.
\end{abstract}

\section{\label{sec:Introduction}Introduction}

Surface representations based on control polyhedra come in various
guises, including recursive subdivision~\cite{Subdivision}, generalized
splines~\cite{Macropatch} and `manifold' approaches~\cite{Manifold}.
In this paper we examine a variation of the manifold-based construction
exemplified by the works of Zorin~\cite{Zorin,Zorin2}, and recently
by Fang~\cite{Fang}.

\begin{figure}
\begin{centering}
\includegraphics[width=.4\textwidth]{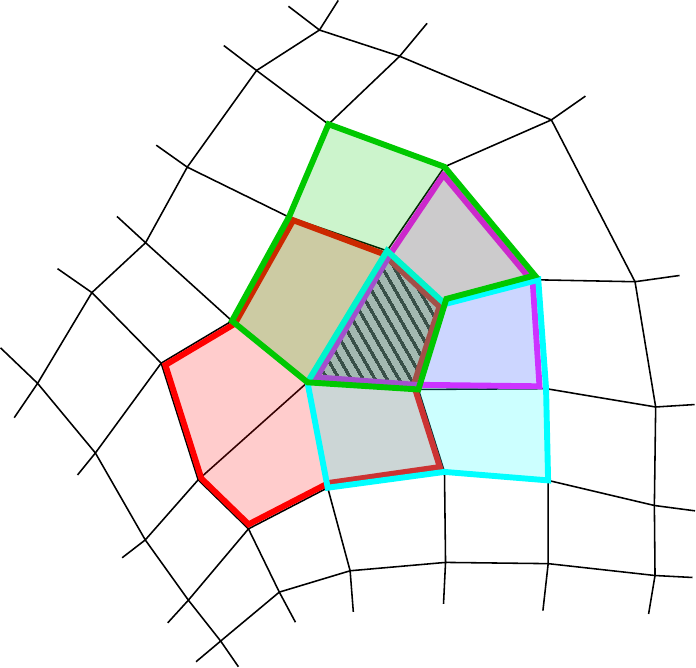}
\par\end{centering}
\caption{\label{fig:Idea}Overlapping quadratic nets around a quad.}

\end{figure}
\begin{figure}
\includegraphics[width=0.19\textwidth]{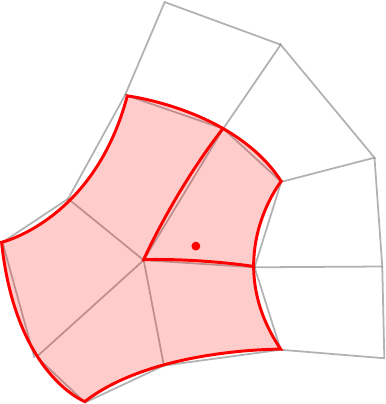}\hfill{}\includegraphics[width=0.19\textwidth]{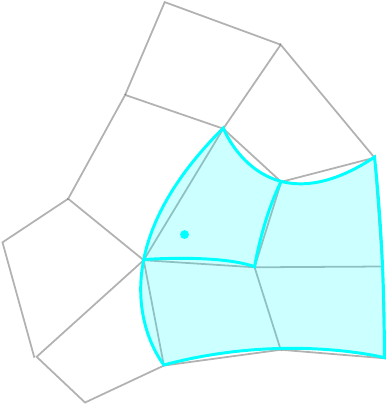}\hfill{}\includegraphics[width=0.19\textwidth]{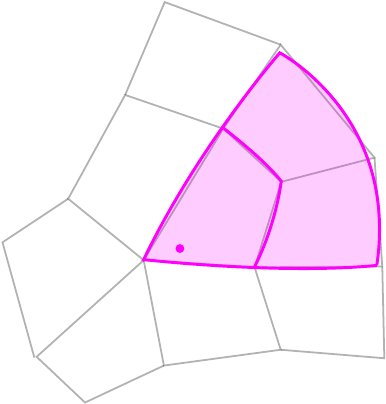}\hfill{}\includegraphics[width=0.19\textwidth]{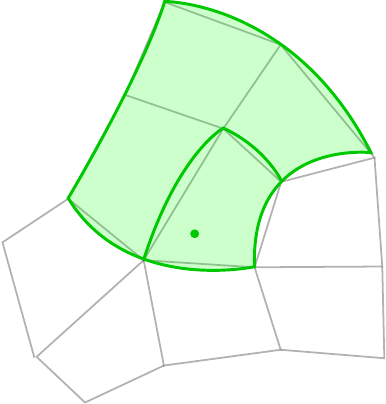}\hfill{}\includegraphics[width=0.19\textwidth]{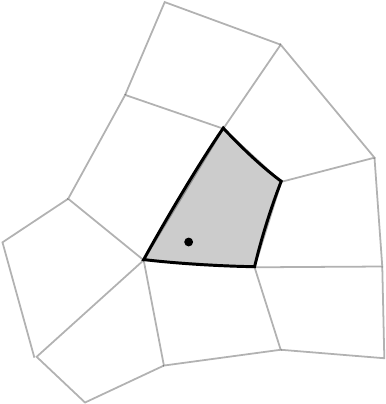}

\caption{\label{fig:Blending}Blending the local interpolants. Dots indicate
a point of evaluation in each; the last image shows the resulting
patch.}

\end{figure}

Figure~\ref{fig:Idea} shows the basic idea. The input is a closed
mesh. We assume that it only contains quads -- if not, we can perform
a central split on all faces (similarly to a Catmull--Clark subdivision
step) to get rid of the multi-sided patches while retaining all original
vertices. For each quad, the 1-rings around its corners define (multi-sided)
quadratic control nets. We can generate four local patches interpolating
these control points, and evaluate them in the parametric region associated
with the quad (Fig.~\ref{fig:Blending}). The resulting surface is
created by blending these together.

Note that while most similar methods are approximative, here we focus
on patches that interpolate the input points. The interpolation criterion
makes the interpolants more similar in the given region, which may
enhance the quality of the resulting surface.

The rest of the paper is organized as follows. Section~\ref{sec:Regular-meshes}
shows the construction in the simple case of regular quadmeshes. Handling
of irregular vertices is explained in Section~\ref{sec:Irregular-vertices},
with the definition of Quadratic Generalized B{\'e}zier (QGB) patches
in Section~\ref{subsec:Quadratic-generalized-B=0000E9zier} and a
parameterization based on rational B{\'e}zier curves in Section~\ref{subsec:Parameterization}.
Some test results are presented in Section~\ref{sec:Results}; notes
on future work conclude the paper.

\section{\label{sec:Regular-meshes}Regular meshes}

In a regular (closed) mesh all vertices have a valency of 4, so all
control nets around a quad will be quadrilateral. We use quadratic
tensor-product B{\'e}zier patches for the interpolants.

Let $\mathbf{C}_{i}$ and $\mathbf{E}_{i}$ ($i=1\dots4$) denote
the corner and edge control points, with $\mathbf{E}_{i}$ being between
$\mathbf{C}_{i-1}$ and $\mathbf{C}_{i}$ (with cyclical indexing).
The middle control point is denoted by $\mathbf{M}$. Then the interpolant
is defined as
\begin{equation}
\mathbf{I}(u,v)=\sum_{i=0}^{2}\sum_{j=0}^{2}\mathbf{P}_{ij}B_{i}^{2}(u)B_{j}^{2}(v),
\end{equation}
where $B_{k}^{d}(t)$ is the $k$-th Bernstein polynomial of degree
$d$ at parameter $t$, and
\begin{align}
\mathbf{P}_{00} & =\mathbf{C}_{3}, & \mathbf{P}_{10} & =\hat{\mathbf{E}}_{4}, & \mathbf{P}_{20} & =\mathbf{C}_{4},\nonumber \\
\mathbf{P}_{01} & =\hat{\mathbf{E}}_{3}, & \mathbf{P}_{11} & =\frac{1}{4}(16\mathbf{M}-\sum_{i=1}^{4}(\mathbf{C}_{i}+2\hat{\mathbf{E}}_{i})), & \mathbf{P}_{21} & =\hat{\mathbf{E}}_{1},\nonumber \\
\mathbf{P}_{02} & =\mathbf{C}_{2}, & \mathbf{P}_{12} & =\hat{\mathbf{E}}_{2}, & \mathbf{P}_{22} & =\mathbf{C}_{1}.
\end{align}
Here $\hat{\mathbf{E}}_{i}=2\mathbf{E}_{i}-\frac{1}{2}(\mathbf{C}_{i-1}+\mathbf{C}_{i})$
is the control point position s.t.~$[\mathbf{C}_{i-1},\hat{\mathbf{E}}_{i},\mathbf{C}_{i}]$
define a quadratic B{\'e}zier curve interpolating $\mathbf{E}_{i}$.

The parameterization of the quad is in $[0,1]^{2}$, and is rotated
locally with the following interpolation constraints in mind ($\mathbf{S}$
denotes the patch to be created):
\begin{align}
\mathbf{S}(0,0) & =\mathbf{M}, & \mathbf{S}(1,0) & =\mathbf{E}_{1},\nonumber \\
\mathbf{S}(0,1) & =\mathbf{E}_{2}, & \mathbf{S}(1,1) & =\mathbf{C}_{1}.\label{eq:param-rot}
\end{align}
The interpolant point corresponding to the $(u,v)$ point in the quad's
domain is then $\left((u+1)/2,(v+1)/2\right)$.

Consequently, the quad patch can be defined as the blend of the four
interpolants:
\begin{equation}
\mathbf{S}(u,v)=\sum_{i=1}^{4}\mathbf{I}_{i}\left(\frac{u_{i}+1}{2},\frac{v_{i}+1}{2}\right)\Phi(u_{i},v_{i}).
\end{equation}
The local parameterizations are defined as
\begin{align}
u_{1} & =u, & v_{1} & =v,\nonumber \\
u_{2} & =v, & v_{2} & =1-u,\nonumber \\
u_{3} & =1-u, & v_{3} & =1-v,\nonumber \\
u_{4} & =1-v, & v_{4} & =u.
\end{align}
The blending function is the product of two 1-argument blends
\begin{equation}
\Phi(u,v)=\Psi(u)\cdot\Psi(v)
\end{equation}
with the following constraints:
\begin{align}
\Psi(0) & =1, & \Psi(1) & =0, & \Psi^{(k)}(0) & =\Psi^{(k)}(1)=0
\end{align}
for some $k>0$. For finite $k$, Hermite blends can be applied:
\begin{equation}
\Psi(t)=\sum_{i=0}^{k}B_{i}^{2k+1}(t).
\end{equation}
For $k=\infty$ there are several options, including bump functions
and expo-rational B-splines~\cite{Blend}. We have used the $k=2$
Hermite blend for all examples in the paper.

\section{\label{sec:Irregular-vertices}Irregular vertices}

An irregular vertex generates a non-quadrilateral control net. We
interpolate these points by quadratic generalized B{\'e}zier patches,
as defined below.

\subsection{\label{subsec:Quadratic-generalized-B=0000E9zier}Quadratic Generalized
B{\'e}zier (QGB) interpolants}

\begin{figure}
\subfloat[\label{fig:Bilinear-map}Bilinear map]{\begin{centering}
\includegraphics[height=0.14\textheight]{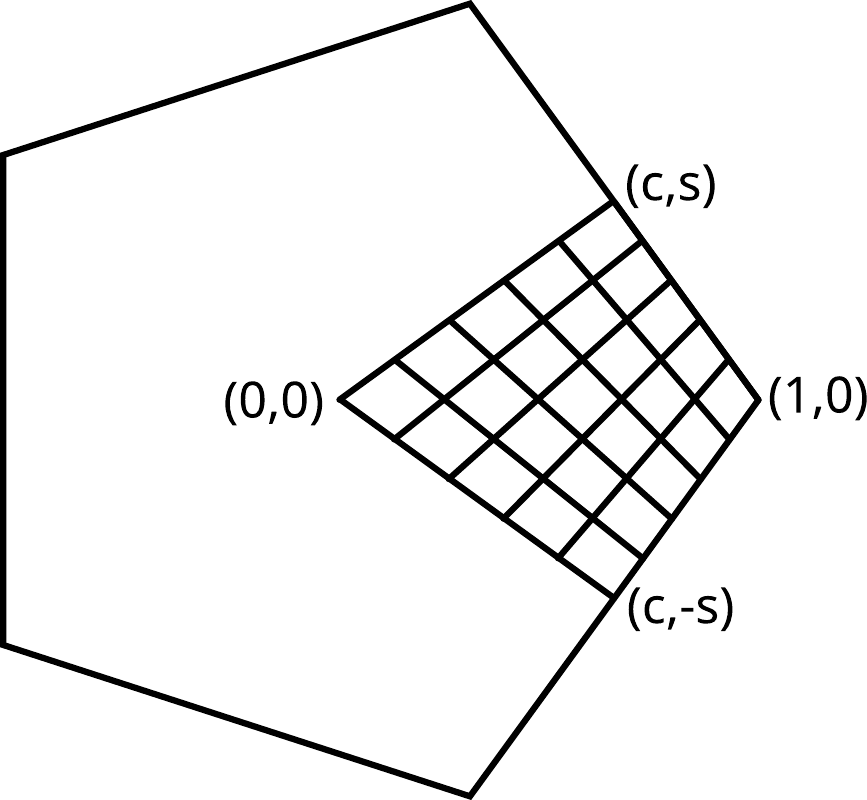}
\par\end{centering}
}\hspace*{\fill}\subfloat[\label{fig:Discontinuity}Discontinuity]{\begin{centering}
\includegraphics[height=0.14\textheight]{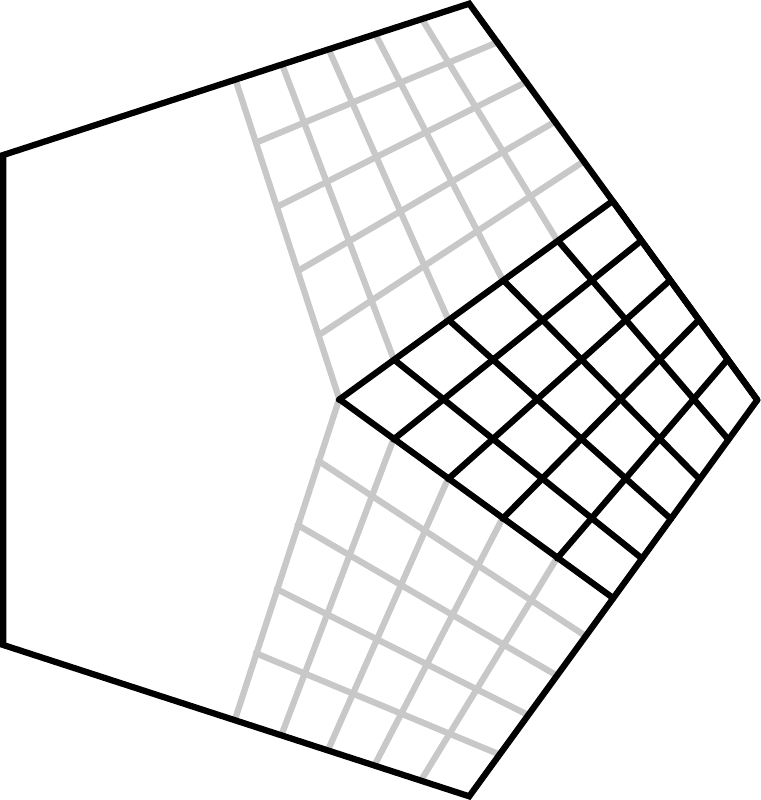}
\par\end{centering}
}\hspace*{\fill}\subfloat[\label{fig:Rational-map}Rational curve map]{\begin{centering}
\includegraphics[height=0.14\textheight]{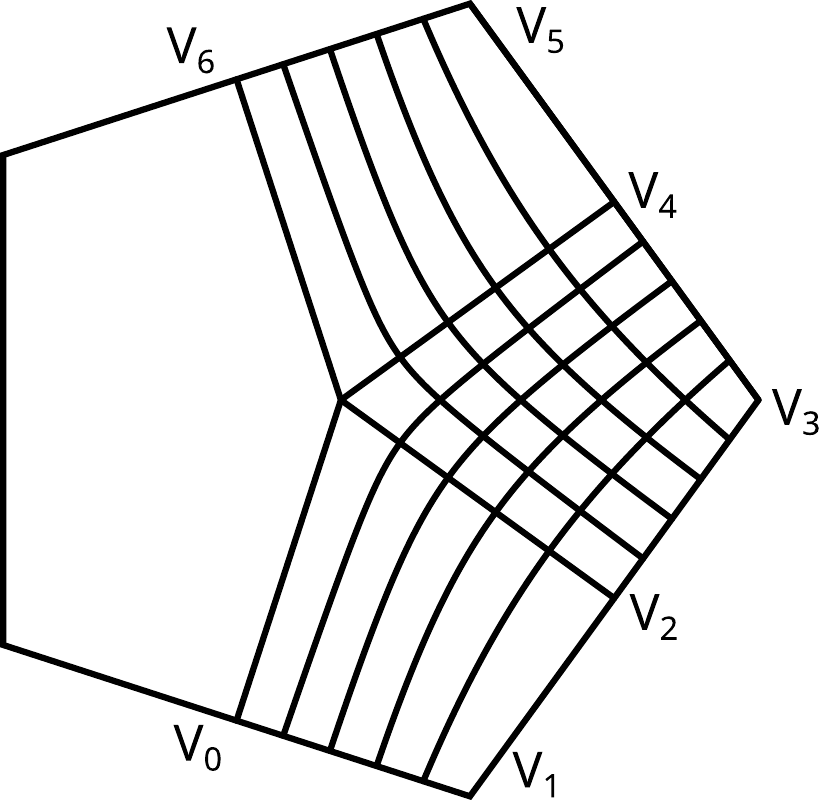}
\par\end{centering}
}\hspace*{\fill}\subfloat[\label{fig:Rational-map3}Triangle case]{\begin{centering}
\includegraphics[height=0.14\textheight]{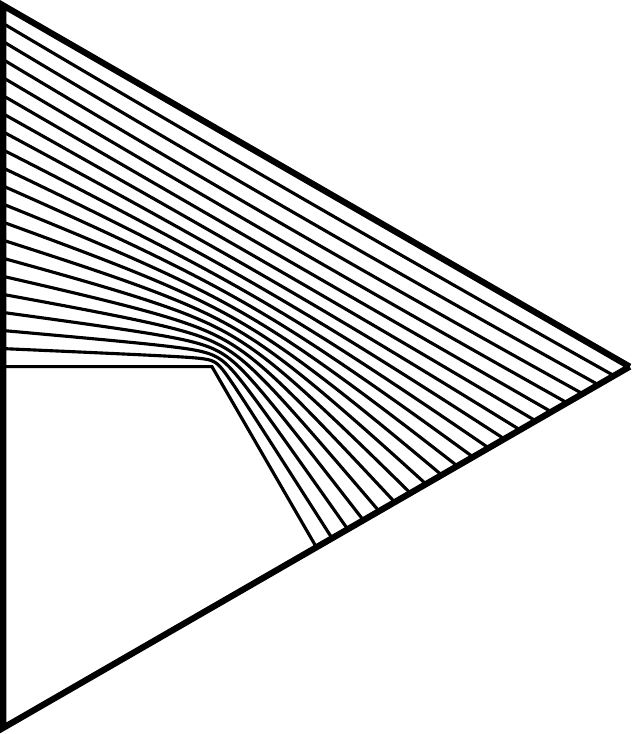}
\par\end{centering}
}

\caption{\label{fig:Mappings}Quad $\rightarrow$ interpolant mappings.}
\end{figure}
Generalized B{\'e}zier (GB) patches~\cite{GB} are normally defined only
for cubic or higher degrees, but is easy to further generalize the
construction to the quadratic case. As this will be a $C^{0}$ interpolant,
its equation will be even simpler than that of the original.

The surface is defined over the regular $n$-sided polygon $\{(\cos\frac{2k\pi}{n},\sin\frac{2k\pi}{n})\}$,
$k=0\dots n-1$, and for each side we create local coordinate mappings
\begin{align}
s_{i}(u,v) & =\lambda_{i}/(\lambda_{i-1}+\lambda_{i}), & d_{i}(u,v) & =1-\lambda_{i-1}-\lambda_{i},
\end{align}
where $\{\lambda_{i}\}$ are the Wachspress coordinates~\cite{Wachspress}
of $(u,v)$ relative to the domain polygon.

Using the notations of the previous section,
\begin{align}
\mathbf{I}(u,v) & =\sum_{i=1}^{n}\left(\mathbf{C}_{i-1}\frac{1}{2}B_{0}^{2}(s_{i})+\hat{\mathbf{E}}_{i}B_{1}^{2}(s_{i})+\mathbf{C}_{i}\frac{1}{2}B_{2}^{2}(s_{i})\right)B_{0}^{2}(d_{i})\nonumber \\
 & \qquad\quad\,+\mathbf{P}_{0}B_{0}(u,v).
\end{align}
Here $B_{0}$ denotes the weight deficiency
\begin{equation}
B_{0}(u,v)=1-\sum_{i=1}^{n}\left(\frac{1}{2}B_{0}^{2}(s_{i})+B_{1}^{2}(s_{i})+\frac{1}{2}B_{2}^{2}(s_{i})\right)B_{0}^{2}(d_{i}),
\end{equation}
and the central control point $\mathbf{P}_{0}$ is defined s.t.~the
patch interpolates $\mathbf{M}$:
\begin{equation}
\mathbf{P}_{0}=\frac{n^{2}\mathbf{M}-\sum_{i=1}^{n}\left(\mathbf{C}_{i}+2\hat{\mathbf{E}}_{i}\right)}{n(n-3)}.
\end{equation}
Note that for $n=4$ this is exactly the same as the tensor-product
interpolant in Section~\ref{sec:Regular-meshes}. It is also easy
to see that for $n=3$ this is a quadratic triangular B{\'e}zier patch.

\subsection{\label{subsec:Parameterization}Parameterization}

We still need a mapping between the quad domain $[0,1]^{2}$ and the
interpolant's domain (a regular polygon inscribed in the unit circle).
The associated part of the latter is a \emph{kite} defined by the
points $(0,0)$, $(c,-s)$, $(1,0)$ and $(c,s)$, where $c=\frac{1}{2}(\cos(2\pi/n)+1)$
and $s=\frac{1}{2}\sin(2\pi/n)$, see Figure~\ref{fig:Mappings}.
A na\"\i ve approach would be to use a bilinear mapping (Fig.~\ref{fig:Bilinear-map}).
This presents a problem however: with this simple mapping adjacent
parts of the same interpolant would be parameterized discontinuously
(Fig.~\ref{fig:Discontinuity}). In the rest of this section, we
will construct a mapping that is $C^{\infty}$-continuous (except
at the origin, which is singular).

The proposed method works by creating \emph{pencils} of quadratic
rational B{\'e}zier curves in the two parametric directions, and the mapping
of $(u,v)$ is defined to be the intersection of the $u$-curve with
the $v$-curve -- see Figure~\ref{fig:Rational-map}, where the
following notations are also shown:
\begin{align}
\mathbf{V}_{0} & =(c+\hat{c},-s-\hat{s}), & \mathbf{V}_{1} & =(2c-1,-2s),\nonumber \\
\mathbf{V}_{2} & =(c,-s), & \mathbf{V}_{3} & =(1,0),\nonumber \\
\mathbf{V}_{4} & =(c,s), & \mathbf{V}_{5} & =(2c-1,2s),\nonumber \\
\mathbf{V}_{6} & =(c+\hat{c},s+\hat{s}),
\end{align}
where $\hat{c}=\frac{1}{2}(\cos(4\pi/n)-1)$ and $\hat{s}=\frac{1}{2}\sin(4\pi/n)$.
The control points and rational weights of $u$-curves are given as
\begin{align}
\mathbf{R}_{0} & =\mathbf{V}_{0}(1-u)+\mathbf{V}_{1}u, & w_{0} & =1,\nonumber \\
\mathbf{R}_{1} & =\mathbf{V}_{2}u\exp(u^{2}-u), & w_{1} & =1/u,\nonumber \\
\mathbf{R}_{2} & =\mathbf{V}_{4}(1-u)+\mathbf{V}_{3}u, & w_{2} & =1.
\end{align}
Similarly for $v$-curves:
\begin{align}
\mathbf{R}_{0} & =\mathbf{V}_{6}(1-v)+\mathbf{V}_{5}v, & w_{0} & =1,\nonumber \\
\mathbf{R}_{1} & =\mathbf{V}_{4}v\exp(v^{2}-v), & w_{1} & =1/v,\nonumber \\
\mathbf{R}_{2} & =\mathbf{V}_{2}(1-v)+\mathbf{V}_{3}v, & w_{2} & =1.
\end{align}
Then the curves themselves can be evaluated with the formula
\begin{equation}
\mathbf{r}(t)=\frac{\sum_{i=0}^{2}\mathbf{R}_{i}w_{i}B_{i}^{2}(t)}{\sum_{i=0}^{2}w_{i}B_{i}^{2}(t)}.
\end{equation}

The exponential term in the middle control point is there to accommodate
for the triangular domain where the $\mathbf{V}_{2}-\mathbf{V}_{3}-\mathbf{V}_{5}-\mathbf{V}_{6}-(0,0)$
polygon is concave, and this helps pulling back the curve near the
origin. Figure~\ref{fig:Rational-map3} shows $v$-curves in a triangle
for $v=k/20$ values, $k=0\dots20$.

For a given $(u,v)$ point in the quad domain, we need to intersect
the corresponding $u$- and $v$-curves. This is easily done with
nested golden section searches in the $[\frac{1}{2},1]$ interval~\cite{GoldenSection}.
While this is a bit expensive, it can (and should) be precomputed.

\subsection{\label{subsec:Triangular-patches}Triangular patches}

\begin{figure}
\begin{centering}
\hspace*{\fill}\includegraphics[width=0.3\textwidth]{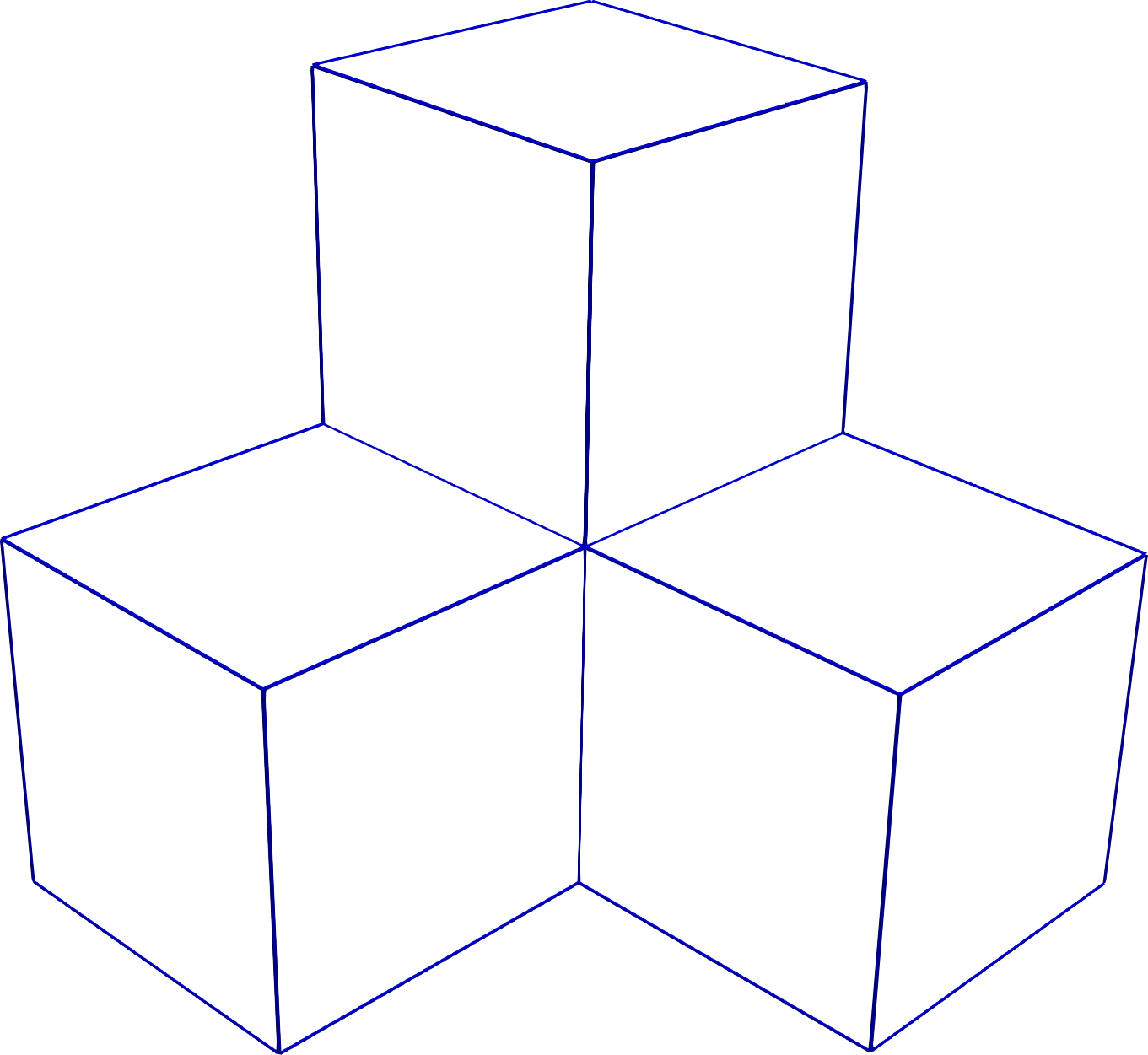}\hspace*{\fill}\includegraphics[width=0.3\textwidth]{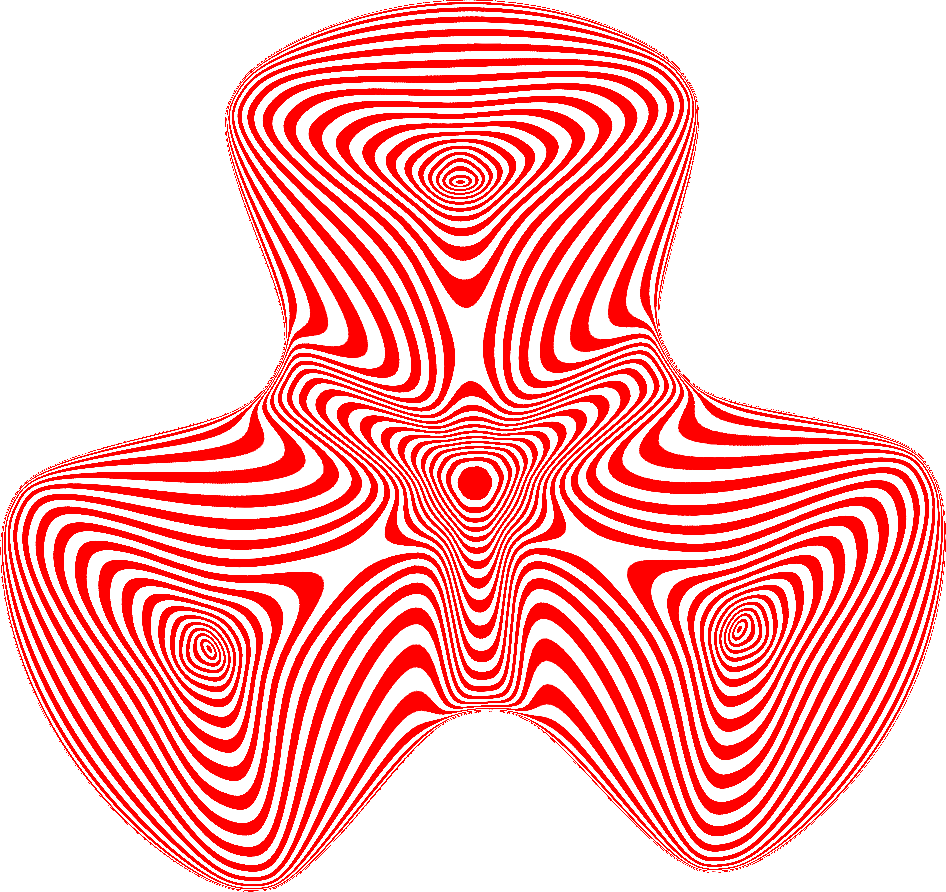}\hspace*{\fill}\includegraphics[width=0.3\textwidth]{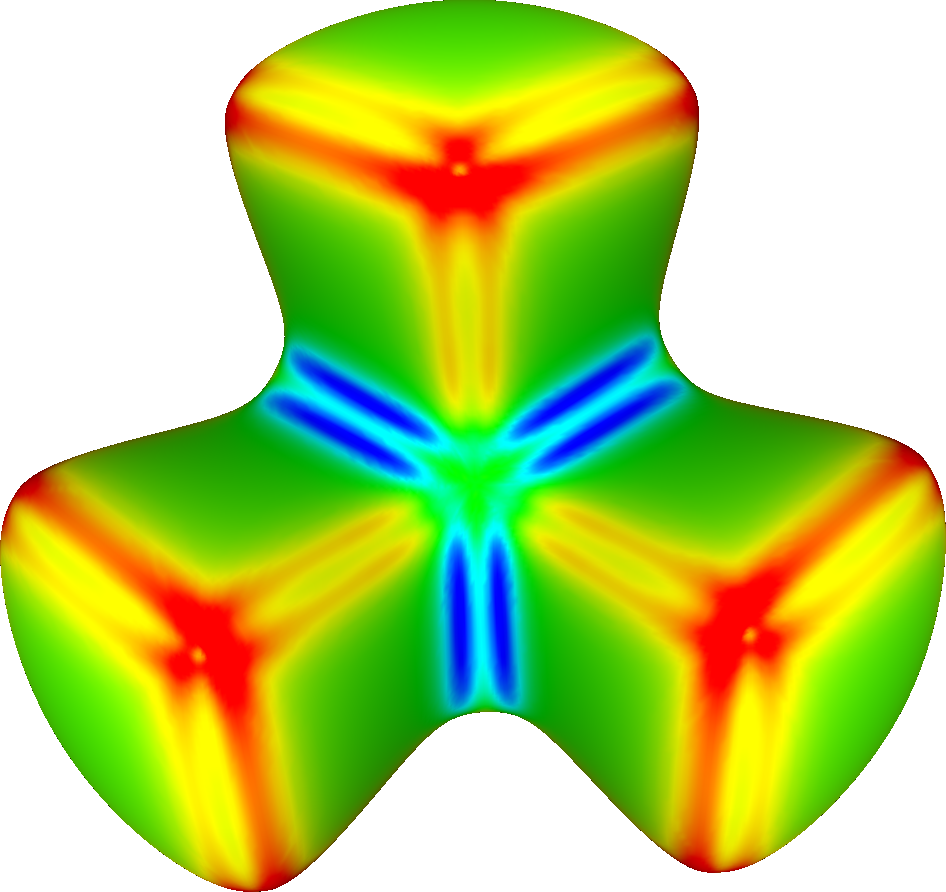}\hspace*{\fill}
\par\end{centering}
\caption{\label{fig:Trebol}Trebol model (left: cage, center: isophotes,
right: mean curvature).}
\end{figure}
The three-sided QGB patch is a quadratic B{\'e}zier triangle, and as such,
does not have a central control point. Since we want to interpolate
the middle point $\mathbf{M}$, we should degree-elevate the boundaries
to cubic and use a cubic B{\'e}zier triangle, which has an extra degree
of freedom:
\begin{align}
\mathbf{P}_{300} & =\mathbf{C}_{1},\quad\mathbf{P}_{030}=\mathbf{C}_{2},\quad\mathbf{P}_{003}=\mathbf{C}_{3},\nonumber \\
\mathbf{P}_{210} & =\frac{1}{3}(\mathbf{C}_{1}+2\hat{\mathbf{E}}_{2}),\quad\mathbf{P}_{120}=\frac{1}{3}(\mathbf{C}_{2}+2\hat{\mathbf{E}}_{2}),\nonumber \\
\mathbf{P}_{021} & =\frac{1}{3}(\mathbf{C}_{2}+2\hat{\mathbf{E}}_{3}),\quad\mathbf{P}_{012}=\frac{1}{3}(\mathbf{C}_{3}+2\hat{\mathbf{E}}_{3}),\nonumber \\
\mathbf{P}_{102} & =\frac{1}{3}(\mathbf{C}_{3}+2\hat{\mathbf{E}}_{1}),\quad\mathbf{P}_{201}=\frac{1}{3}(\mathbf{C}_{1}+2\hat{\mathbf{E}}_{1}),
\end{align}
and
\begin{equation}
\mathbf{P}_{111}=\frac{1}{6}(27\mathbf{M}-\sum_{\max(i,j,k)=3}\mathbf{P}_{ijk}-3\sum_{\max(i,j,k)=2}\mathbf{P}_{ijk}).
\end{equation}
Then the patch can be evaluated by
\begin{equation}
\mathbf{I}(u,v)=\sum_{i+j+k=3}\mathbf{P}_{ijk}\frac{6}{i!j!k!}\lambda_{1}^{i}\lambda_{2}^{j}\lambda_{3}^{k}.
\end{equation}

\subsection{\label{subsec:Alternative-representations}Alternative representations}

There are many multi-sided surface representations that we could use;
for a survey, see our upcoming paper~\cite{Survey}. In particular,
Midpoint and Midpoint Coons patches~\cite{Midpoint} are very well
suited for this task. Our experiments have shown, however, little
difference in the resulting models, while computationally QGB patches
are simpler and more efficient.

\section{\label{sec:Results}Results}

\begin{figure}
\begin{centering}
\hfill
\includegraphics[width=0.4\textwidth]{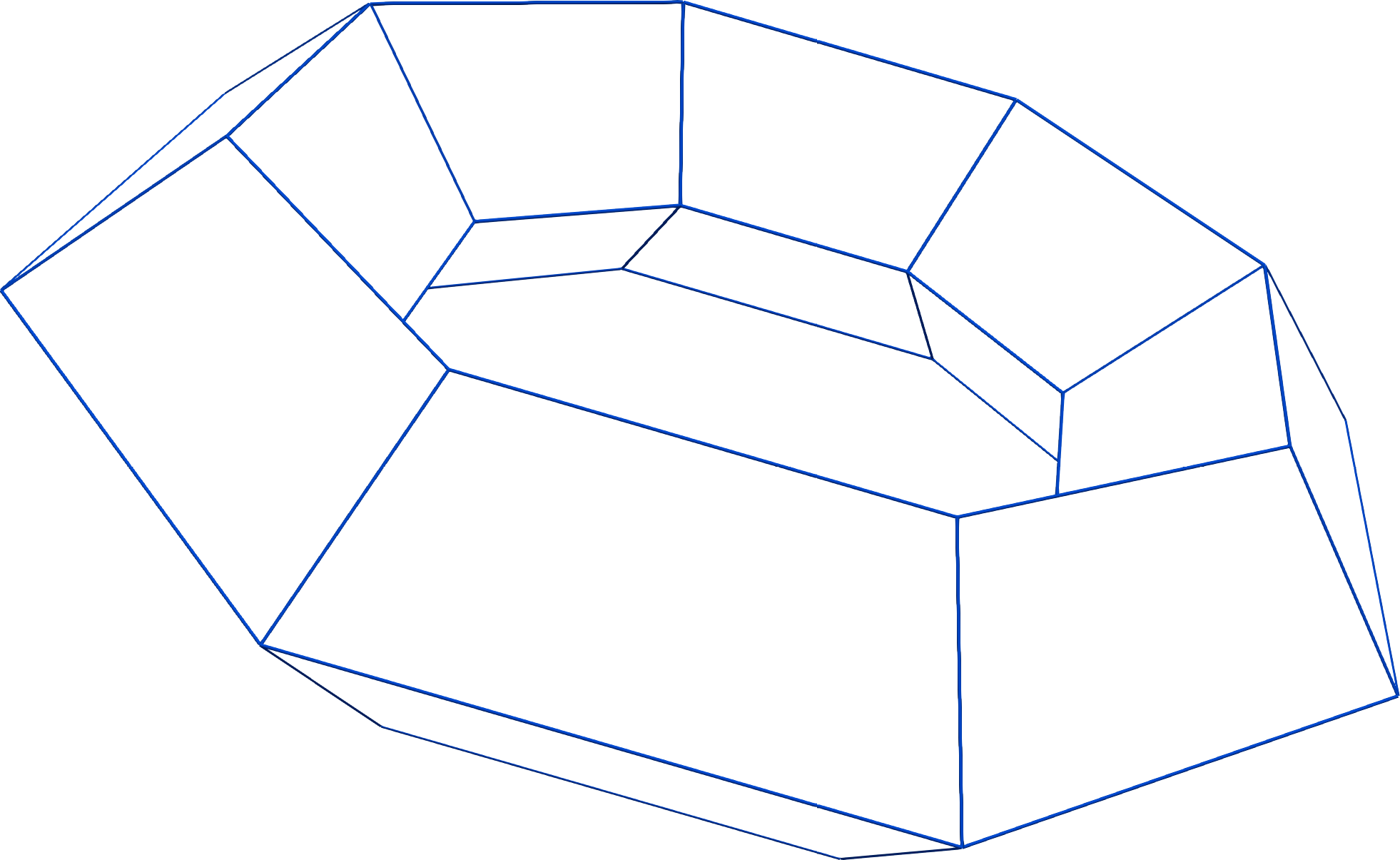}
\hfill
\includegraphics[width=0.4\textwidth]{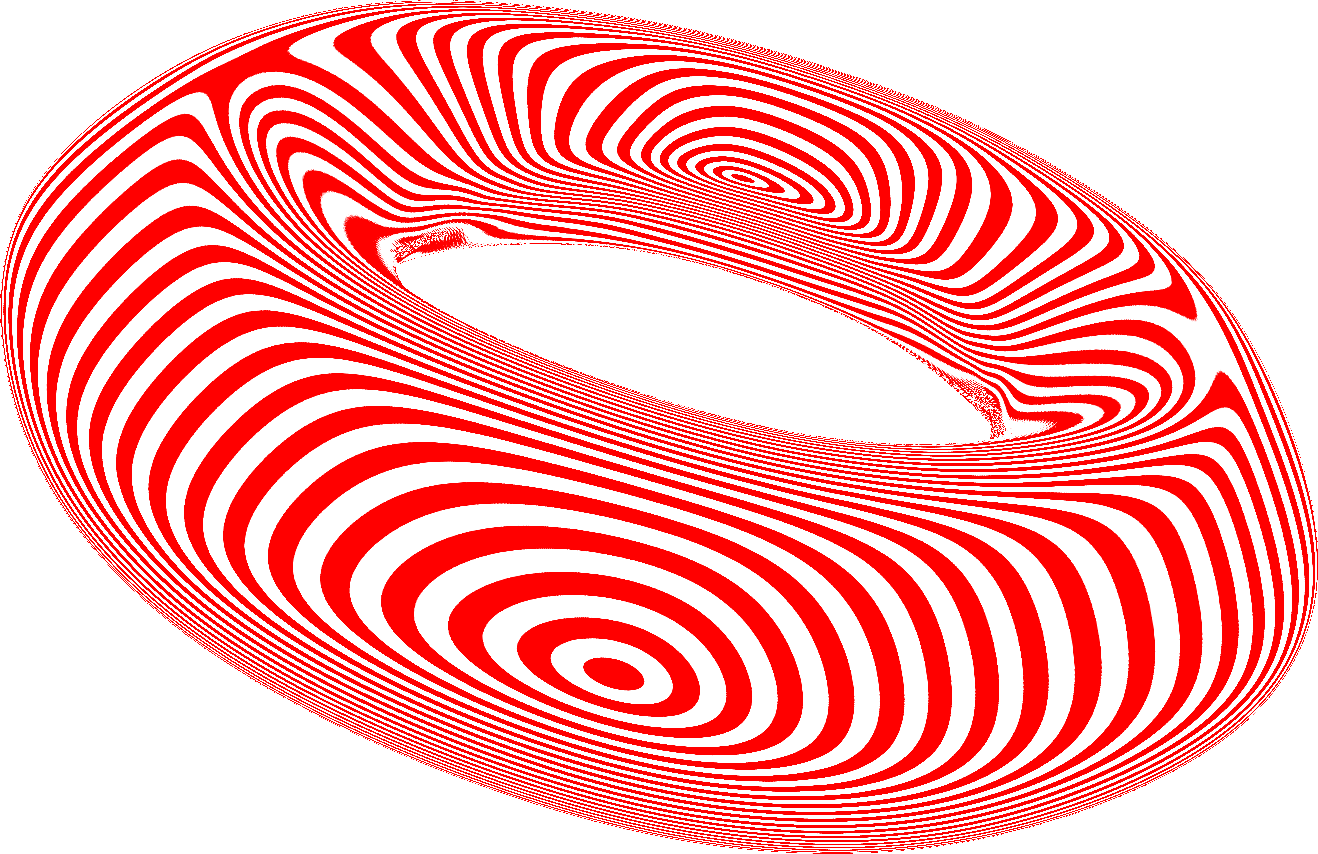}
\hfill
\end{centering}
\caption{\label{fig:Torus}Torus model (left: cage, right: isophote map).}
\end{figure}

\begin{figure}
\begin{centering}
\hspace*{\fill}\includegraphics[width=0.3\textwidth]{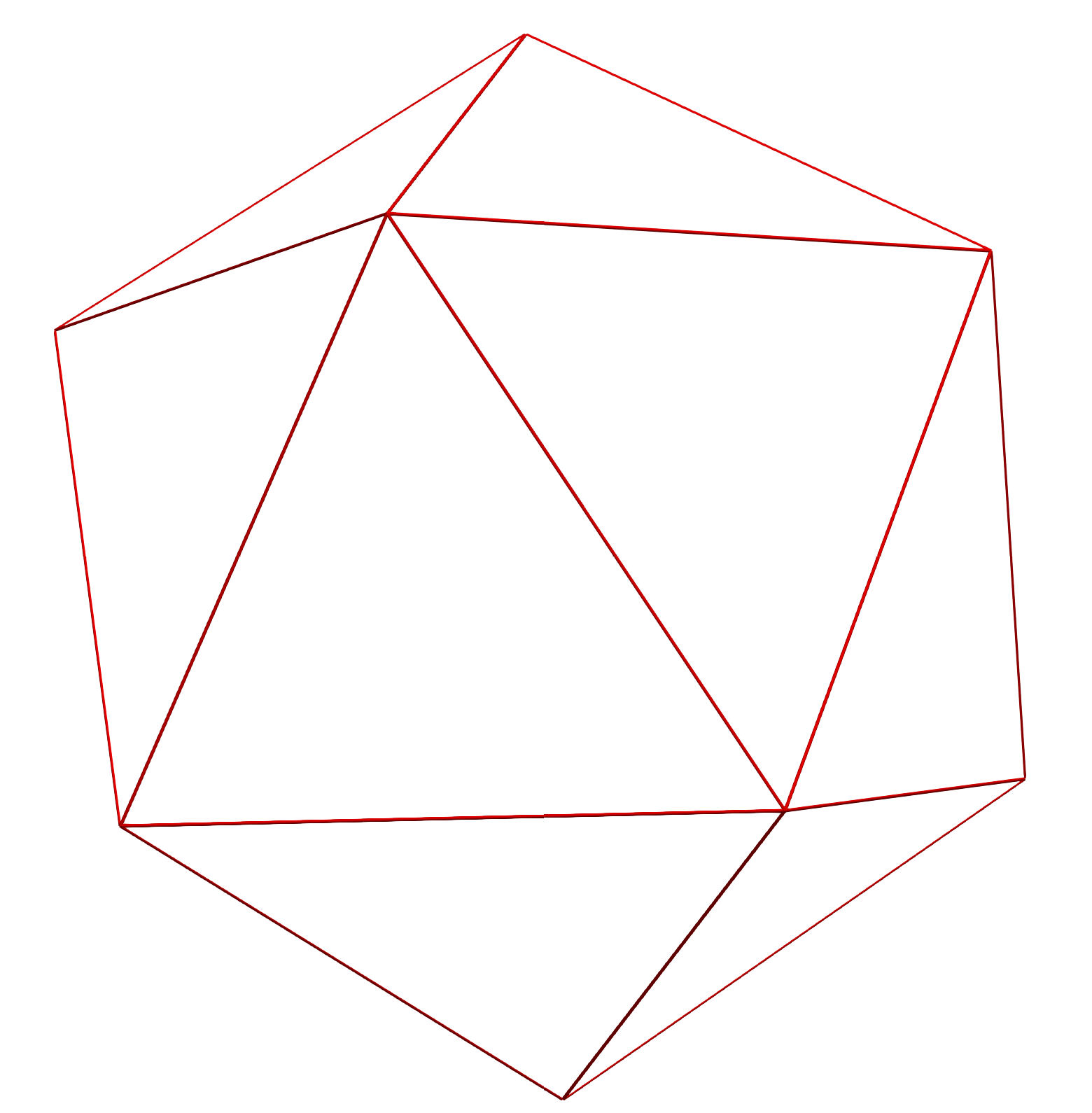}\hspace*{\fill}\includegraphics[width=0.3\textwidth]{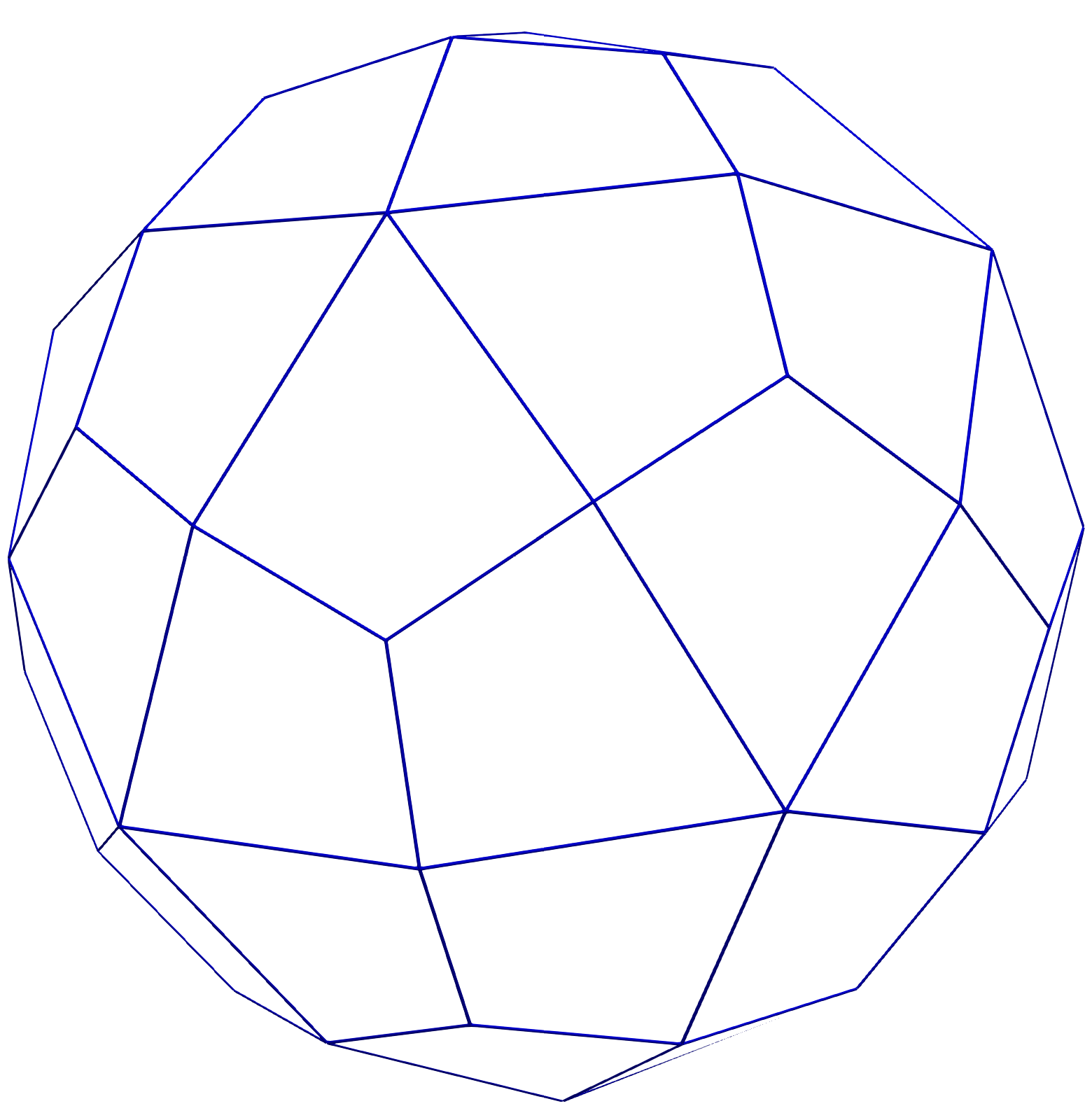}\hspace*{\fill}\includegraphics[width=0.3\textwidth]{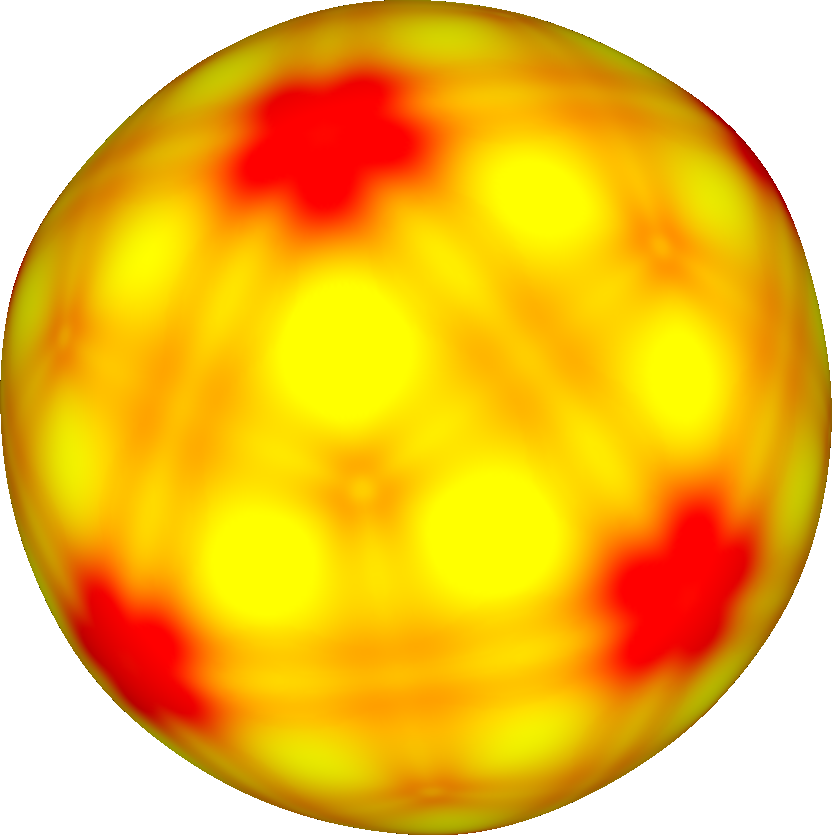}\hspace*{\fill}
\par\end{centering}
\caption{\label{fig:Icosahedron}Icosahedron model (from left to right: triangular
cage, cage after subdivision, mean curvature map).}
\end{figure}

Our first model is a regular quadmesh in the shape of a torus, see
Figure~\ref{fig:Torus}. The result is far from a torus, but the
isophote lines flow smoothly on the surface.

The \emph{trebol} object shown in Figure~\ref{fig:Trebol} is a commonly
used test model as it has vertices of valences 3, 4, 5 and 6. The
isophote lines indicate that the multi-sided patches are of good quality
and connect continuously, while the mean map shows that edges tend
to be a little flat.

The icosahedron model in Figure~\ref{fig:Icosahedron} is composed
of triangles. As a first step, a Catmull--Clark step is performed,
resulting in a quadmesh with 3-, 4- and 5-valent vertices. The mean
curvature map suggests larger values near 5-valent vertices, but otherwise
there are only small fluctuations, corresponding to the mesh edges.

\begin{figure}[t]
\hfill
\subfloat[Cage]{\includegraphics[width=.45\textwidth]{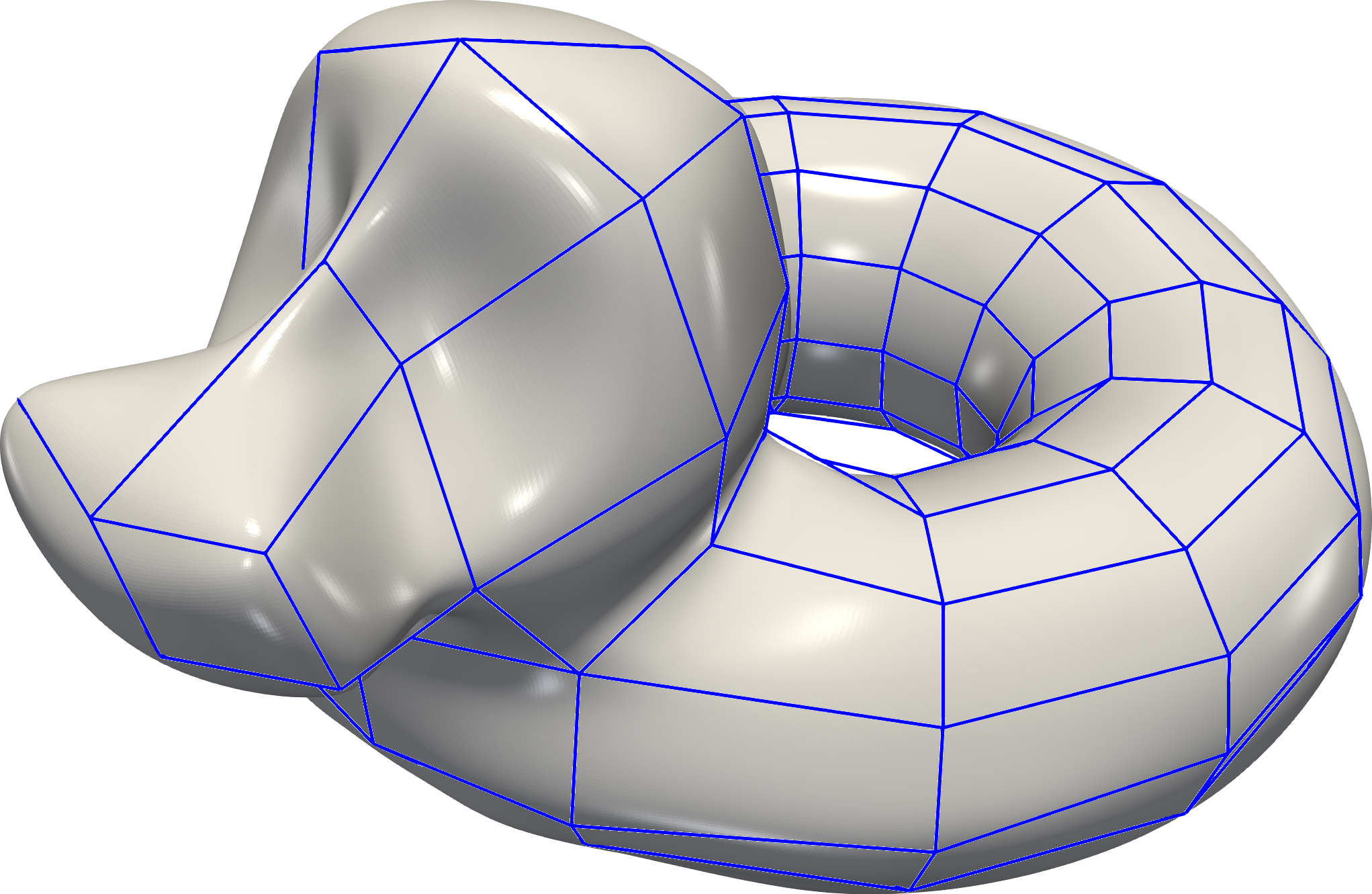}}
\hfill
\subfloat[Contouring]{\includegraphics[width=.45\textwidth]{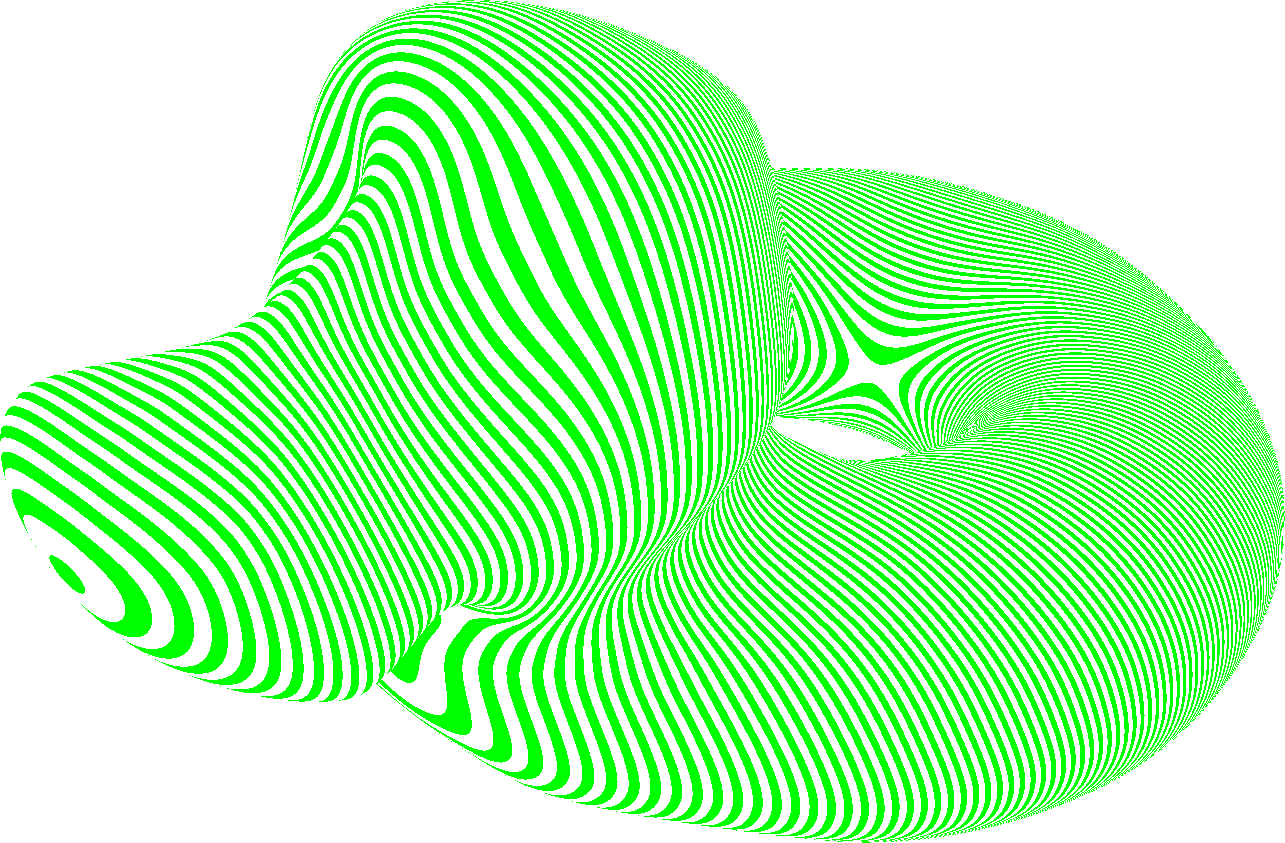}}
\hfill

\hfill
\subfloat[Reflections]{\includegraphics[width=.45\textwidth]{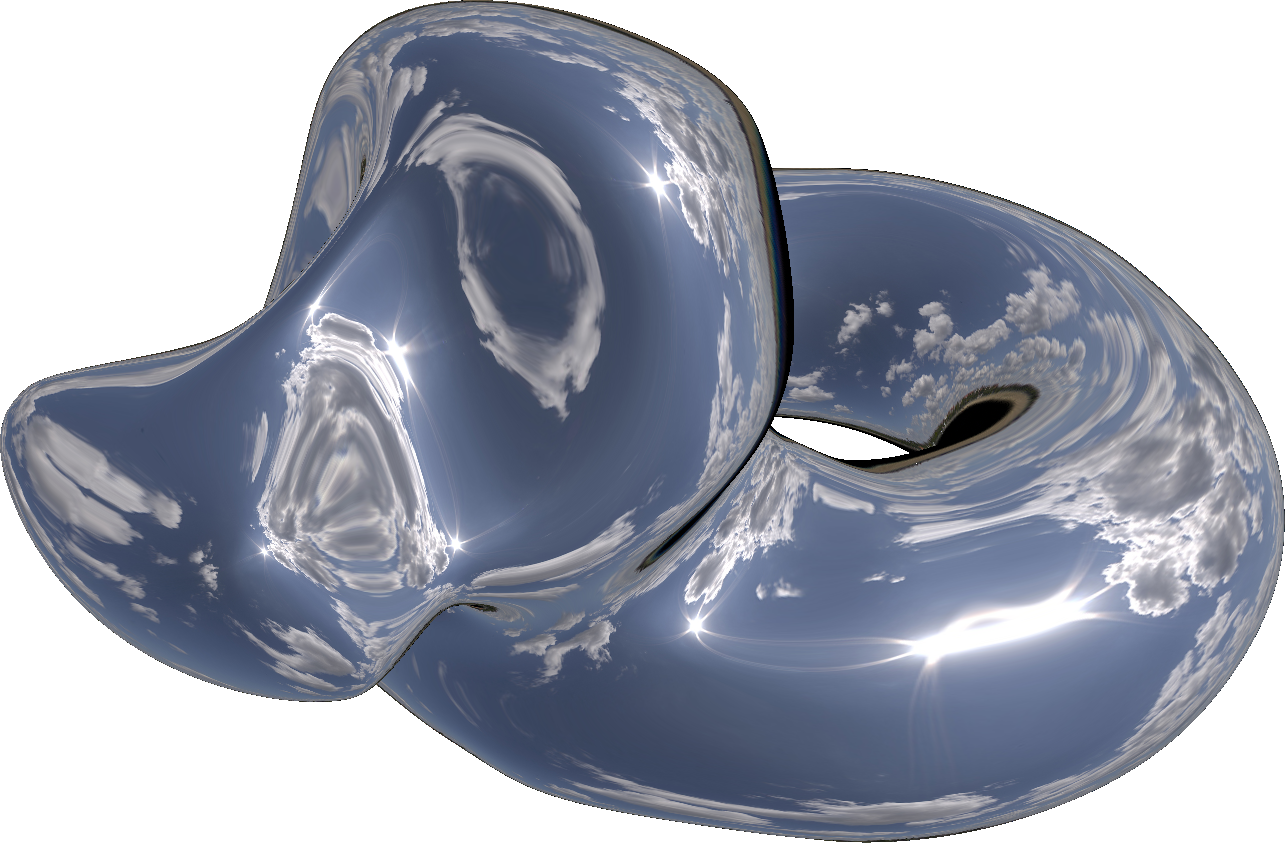}}
\hfill
\subfloat[Mean curvature map]{\includegraphics[width=.45\textwidth]{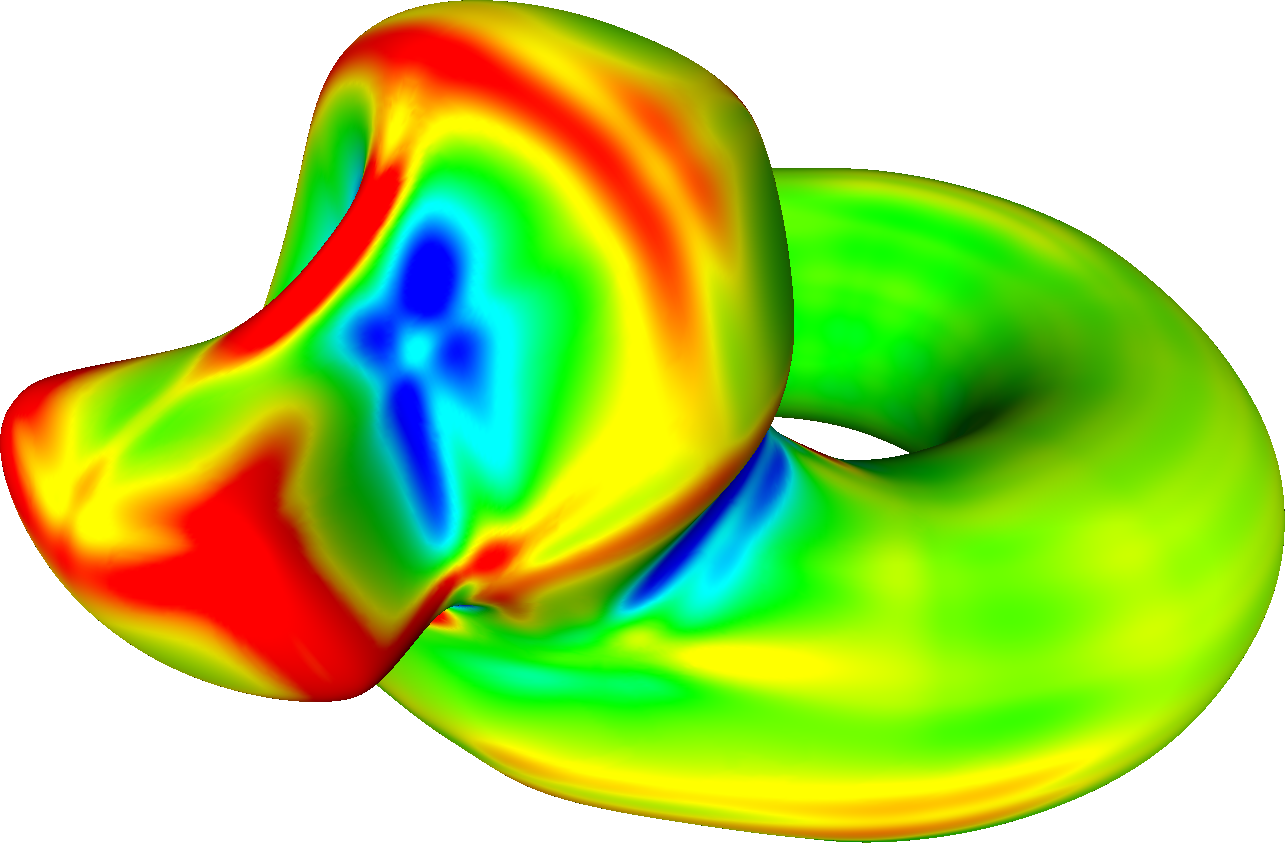}}
\hfill

\caption{\label{fig:Bob}`Bob' model.}
\end{figure}
Finally, Figure~\ref{fig:Bob} shows a more complex
model\footnote{Taken from the code supplied to the paper on K-surfaces~\cite{Bob}.}
with environment mapping, contouring and mean curvature.

\section*{Conclusion and future work}

We have proposed a new construction for a piecewise parametric surface
interpolating a mesh of arbitrary topology. There are many avenues
for further research. Currently our method works only for closed meshes;
it should be generalized to meshes with boundary. Then continuous
boundary constraints (positional and cross-derivative interpolation)
may also be added. Extension with shape parameters can also be considered.
Better quality may be achieved if normal vector interpolation is incorporated.

\section*{Acknowledgments}

This project has been supported by the Hungarian Scientific Research
Fund (OTKA, No.~145970).

\bibliographystyle{plain}
\bibliography{cikkek}

\end{document}